\documentclass[prl,showpacs,twocolumn,amsmath]{revtex4}

\usepackage{graphicx}
\usepackage{bm}

\begin{document}

\title{Additivity of decoherence measures for multiqubit quantum
systems}

\author{Leonid Fedichkin}
\author{Arkady Fedorov}
\author{Vladimir Privman}
\affiliation{Center for Quantum Device Technology, Department of
Physics,\\ Clarkson University, Potsdam, NY 13699}

\begin{abstract}
We introduce new measures of decoherence appropriate for
evaluation of quantum computing
designs. Environment-induced deviation of a quantum system's
evolution from controlled dynamics is quantified by a single
numerical measure. This measure is defined as a maximal norm of
the density matrix deviation. We establish the property of
\emph{additivity\/}: in the regime of the onset of decoherence,
the sum of the individual qubit error measures provides an
estimate of the error for a several-qubit system. This property is
illustrated by exact calculations for a spin-boson model.
\end{abstract}


\date{\today}

\maketitle

Dynamics of open quantum systems has long been a subject of study
in diverse fields \cite{open,vanKampen}. Recent interest in
quantum computing has focused attention
\cite{nonMarkov,short,Privman} on quantifying environmental
effects that cause small deviations from the isolated-system
quantum dynamics. During short time intervals of ``quantum-gate''
functions, environment-induced relaxation/decoherence effects must
be kept below a certain threshold in order to allow fault-tolerant
quantum error correction \cite{qec}. The reduced density matrix of
the quantum system, with the environment traced out, is usually
evaluated within some approximation scheme, e.g., \cite{apscheme}.
In this work, we introduce a new, \emph{additive\/} measure of
the deviation from the ``ideal'' density matrix of an isolated
system. For a single two-state system (a qubit) this measure is
calculated explicitly for the environment modeled as a bath of
harmonic modes \cite{Caldeira}, e.g., phonons. Furthermore, we
establish that for a several-qubit system, the introduced measure
of decoherence is approximately additive for short times. It can
be estimated by summing up the deviation measures of the
constituent qubits, without the need to carry out a many-body
calculation, similar to the approximate additivity expected for
relaxation rates of exponential approach to equilibrium at large
times.

In most quantum computing proposals, quantum algorithms are
implemented in the following way. Evolution of the qubits is
governed by a Hamiltonian consisting of single-qubit operators
and of two-qubit interaction terms \cite{Lloyd,Barenco}. Some
parameters of the Hamiltonian can be controlled externally to
implement the desired algorithm. During each cycle of the
computation the Hamiltonian remains constant. This ideal model
does not include the influence of the environment on the
computation, which necessitates quantum error correction. The
latter involves non-unitary operations \cite{qec} and cannot be
described as Hamiltonian-governed dynamics.

The accepted approach to evaluate environmentally induced
decoherence involves a model in which each qubit is coupled to a
bath of environmental modes \cite{Caldeira}. The reduced density
matrix of the system, with the bath modes traced out, then
describes the time-dependence of the system
dynamics \cite{nonMarkov,Markov,apscheme}. Because of the
interaction with environment, after each cycle the state of the
qubits will be slightly different from the ideal. The resulting
error accumulates at each cycle, so that large-scale quantum
computation is not possible without implementing fault-tolerant
error correction schemes \cite{qec,Kitaev}. These schemes require
the environmentally induced decoherence of the quantum state in
one cycle to be below some threshold. Its value, defined for
uncorrelated single qubit error rates, was estimated \cite{PreskillDiVincenzo} to be between
$10^{-6}$ and $10^{-4}$.

To evaluate error rate for a given system, one should first obtain
the evolution of its density operator. In realistic cases
various approximations are
utilized \cite{nonMarkov,short,Markov,apscheme}. The most familiar are
the Markovian-type approximations \cite{Markov} used to evaluate
approach to the thermal state at large times. It has been pointed
out recently that these approximations are not suitable for
quantum computing purposes because they are usually not valid \cite{vanKampen,short} at low temperatures and for the short cycle times of
quantum computation. Several non-Markovian
approaches \cite{nonMarkov,short,Privman,apscheme} have been developed to evaluate the
short-time dynamics of open quantum systems.

Additional issues arise when one tries to study decoherence of
several-qubit systems.  One has to consider the degree to which
noisy environments of different qubits are
correlated \cite{Storcz}. If all the qubits are effectively
immersed in the same bath, then there is a way to reduce
decoherence
for this group of qubits without error correction procedures, by
encoding the state of one logical qubit in a decoherence free
subspace of the states of several physical qubits
\cite{Palma,DFS}. In
a large scale quantum computer consisting of thousands of qubits,
it is more appropriate to consider qubits immersed in distinct
baths, because these errors represent the ``worst case scenario''
that necessitates error-correction.

Once we have the density matrix $\rho(t)$ for a single- or
few-qubit system evaluated in some approximation, we have to
compare it to the ideal density matrix $\rho^{(i)}(t)$
corresponding to quantum algorithm without environmental
influences. We have to define some measure of decoherence, $D(t)$,
to compare with the fault-tolerance criteria, which are not
specific with regards to the error-rate definition \cite{qec}. It
is convenient to have the measure nonnegative, and vanishing if
and
only if the system evolves in complete isolation. Since explicit
calculations beyond one or very few qubits are exceedingly
difficult, it would be also useful to find a measure which is
\emph{additive\/} (or at least sub-additive), i.e., the measure of
decoherence of a composite system will be the sum (or not
greater than the sum) of the measures of decoherence of its
subsystems. In this work, we identify two measures of
decoherence which have these desirable properties. We prove the
asymptotic additivity and derive explicit results for decoherence
in a short-time approximation.

Typically, the Hamiltonian of an open quantum system interacting
with environment has the form $  H =   H_S  +   H_B  +   H_I
$, where $H_S$ is the internal Hamiltonian, $H_B$ is the
Hamiltonian of the environment (bath), $H_I$ is the system-bath
interaction
term. Over gate-function cycles, and between them, the terms in $H$ will be considered constant. Therefore, the overall density matrix $R(t)$ of the system and bath
evolves according to
$R(t)=\exp{\left(-
iHt/\hbar\right)}R(0)\exp{\left(iHt/\hbar\right)}$.
Usually the initial density matrix $R(0)=\rho(0){\raise2pt\hbox{$\scriptscriptstyle{\otimes}$}} \Theta$ is
assumed
\cite{Markov}
to be a direct product of the initial density matrix of the
system, $\rho(0)$, and the thermalized density matrix of the bath,
$\Theta$. The
reduced density matrix
of the system is obtained by tracing out the bath modes,
$\rho(t)={\rm Tr_B} \, R(t) $.

To quantify the effect of the interaction with the bath
\cite{norm}, we consider the deviation, $\sigma(t)$, of the
reduced density
matrix from the ideal $\rho^{(i)}(t) \equiv
\exp{\left(-iH_S t/\hbar\right)}\rho(0)\exp{\left(iH_S
t/\hbar\right)}$,
\begin{equation}
\sigma(t)  =   \rho(t)  - \rho^{(i)}(t).
\end{equation}
As a numerical measure of the deviation we will use the operator
norm $\left\|\sigma \right\|_{\lambda}$. It is defined \cite{Kato}
as the maximal absolute value of the eigenvalues of the matrix
$\sigma$. For a $2 \times 2$ deviation matrix, $ \left\|   \sigma
\right\|_{\lambda} = ( \left| {\sigma _{11} } \right|^2 + \left|
{\sigma_{12} } \right|^2 )^{1/2}$.

Note that $\left\| \sigma \right\|_{\lambda}$ is not only a
function of time, but it also depends on the initial density
operator
$\rho(0)$. Due to decoherence, it will increase from zero as time
goes by.
However, its time-dependence will also contain oscillations at the
frequencies
of the
internal system dynamics, as illustrated in Fig.\ 1. The effect
of the bath can be better quantified by the \emph{maximal\/}
deviation
norm,
\begin{equation}\label{normD}
  D(t) = \sup_{\rho (0)}\big(\| \sigma (t,\rho (0))\|_{\lambda}
  \big).
\end{equation}
One can show that $0 \leq D(t) \leq 1$.  This measure of
decoherence will typically increase monotonically from zero at
$t=0$, saturating at large times at a value
$D(\infty) \leq 1$.
The definition of the maximal decoherence measure $D(t)$ looks
rather complicated for a general multiqubit system. However, in
what follows we show that it can be evaluated in closed
form for short times, appropriate for quantum computing,
for a single-qubit (two-state) system. We then establish an
approximate additivity that allows us to estimate $D(t)$ for
several-qubit systems as well.
\begin{figure}
\includegraphics[width=8cm, height=8cm]{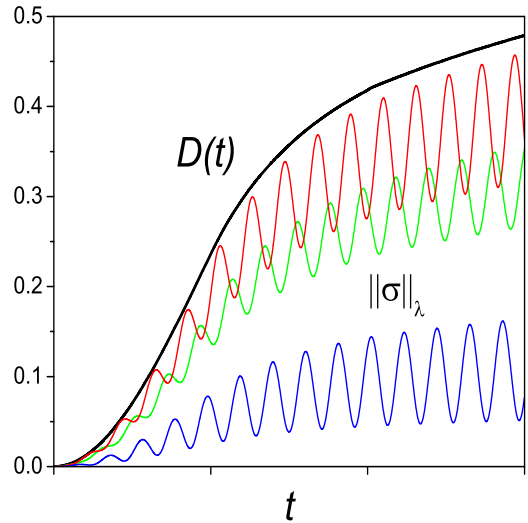}
\caption{\label{fig:3} Schematic plot of the deviation norm for
a two-level system interacting with the
Ohmic bath \cite{open} of bosonic modes in the short-time approximation \cite{short}. The oscillatory curves illustrate the
$\left\| \sigma \right\|_{\lambda}$ norms for three different initial
states $\rho (0)$. The monotonic curve shows the $D$ norm, obtained as $\left\|
\sigma \right\|_{\lambda}$ maximized (\ref{normD}) over $\rho (0)$. Note that in this case $D(\infty)=1/2$.} \label{fig3}
\end{figure}

To be specific, let us outline the calculation of this measure
of decoherence of a two-level system (spin, qubit) interacting
with a bosonic bath of modes. Thus, we take
\begin{eqnarray}\label{1}
 H_S&=&{-(\hbar\Omega/2)}\sigma_z ,\quad H_B= \hbox{$\sum_k
$}{\hbar\omega_k
a_k^{\dagger}
 a_k^{\vphantom{\dagger}}},\\
H_I&=&\sigma _x \hbox{$\sum_k$} {(g_k^{\vphantom{\dagger}} a_k^{\dagger}+ g_k^{*\vphantom{\dagger}} a_k^{\vphantom{\dagger}}
)}.\label{3}
\end{eqnarray}
Here $\omega_k$ are the bath mode frequencies, $a_k^{\vphantom{\dagger}},\,
a_k^{\dagger}$ are the bosonic annihilation and creation
operators,
$g_k$ are the coupling constants, $ \sigma_x$ and $\sigma_z$ are
the Pauli matrices, and $\hbar\Omega>0$ is the energy gap between
the ground (up, $|1\rangle$) and excited (down, $|2\rangle$)
states of the
qubit.

The dynamics of the system can be obtained in closed form
in the short-time approximation \cite{short}. The expression for
the
density operator
of a spin-1/2 system is
\begin{eqnarray}\label{short-time}
\rho _{mn} (t)\! &=&\!\! \hbox{$\sum\limits_{ { p, q=1,2 \atop \mu
, \nu = \pm
1 }}$} {}  \rho _{pq} (0) \langle m|\mu \rangle\langle \mu
|p\rangle\langle q|\nu
\rangle\langle \nu |n \rangle\nonumber\\
& \times & e^{i[ ( E_q + E_n  - E_p  - E_m )t/2 - B^2 (t)(\eta _\mu -
\eta
_\nu )^2 /4 ]}.
\end{eqnarray}
Here the Roman-labeled states, $|j\rangle$, with $j = m,n,p,q = 1,
2$, are the
eigenstates of $ H_S$ corresponding to the
 eigenvalues $E_j/\hbar=(-1)^j \Omega /2$.
The Greek-labeled states, $|\zeta \rangle$, are the
eigenstates of $\sigma_x$, with the eigenvalues $\eta _\zeta =
\zeta$,
where $\zeta = \mu, \nu = \pm 1$. The spectral
function $B^2(t)$ is defined according to
\begin{equation}\label{spectral function}
 B^2 (t) = 8 \hbox{$\sum_k$} {}  {{\left | {g_k }
\right | ^2 }}{{\omega _k ^{-2} }}\sin ^2 ({{{\omega _k
t}/{2}}})\coth ({{{\beta \omega _k } /
 2)}}.
\end{equation}
Here $\beta=\hbar/k_B T$, $k_B$ is the Boltzmann constant, and $T$
is the initial bath temperature. This function has been
studied in \cite{Palma,vanKampen,basis}. With
$\rho_{12}(0)=|\rho_{12}(0)|e^{i
\phi}$, we get
\begin{eqnarray}\label{n1qubit}
\left\| \sigma(t)  \right\|_{\lambda}&=& \hbox{$\frac{1}{2}$}\big[
1 - e^{
- B^2 (t)}\big] \big\{ [\rho _{11} (0) - \rho _{22}
(0)]^2 \nonumber \\
 & + & 4 \left| \rho _{12} (0) \right|^2 \sin ^2 [(\Omega
/2)t +\phi] \big\} ^{1/2}.
\end{eqnarray}
In Fig.\ 1, we show schematically the behavior of $ \left\|
\sigma(t)  \right\|_{\lambda} $  for three representative choices
of the initial density matrix $ \rho (0)$.
Generally, $ \left\| \sigma(t)  \right\|_{\lambda} $ increases
with time, reflecting the decoherence of the system. However,
oscillations at the system's internal frequency $\Omega$ are
superimposed, as seen explicitly in (\ref{n1qubit}).

The effect of the bath is better quantified by the maximal
operator norm $D(t)$. In order to maximize over $\rho(0)$, we can
parameterize $\rho(0)=U \left( P |1\rangle\langle
1| + (1 - P) |2\rangle\langle 2| \right)U^{\dagger}$, where $0\leq
P \leq 1$, and $U$ is an arbitrary $2 \times 2$ unitary matrix,
\begin{equation}
U=\left(
\begin{array}{cc}
  e^{i(\alpha+\gamma)}\cos\theta & e^{i(\alpha-\gamma)}\sin\theta
\\
  -e^{i(\gamma-\alpha)}\sin\theta & e^{-
i(\alpha+\gamma)}\cos\theta \\
\end{array}\right) .
\end{equation}
By using the rigorous definition \cite{Kato} of the norm, $\left\| A
\right\|_{\lambda}^2 = \mathop {\sup }_{\psi \neq 0} ( \langle
\psi |A^ \dagger
A|\psi  \rangle
 / \langle \psi
|\psi  \rangle )$, one can show that it suffices to put $P=1$ and
search for the
maximum over the remaining three real parameters $\alpha$,
$\gamma$
and $\theta$. The result, sketched in Fig.\ 1, is
\begin{equation}\label{D1qubit}
D (t)  =\hbox{$\frac{1}{2}$} \big[ 1 - e^{ - B^2 (t)}\big].
\end{equation}

In quantum computing, the error rates can be significantly reduced
by using several physical qubits to encode each logical qubit
\cite{DFS}. Therefore, even before active quantum error correction
is incorporated \cite{qec}, evaluation of decoherence of several
qubits is an important, but formidable task. Consider a system consisting of two
initially \emph{unentangled\/} subsystems $S_1$ and $S_2$, with
decoherence norms $D_{S_1}$ and $D_{S_2}$, respectively. We denote
the density matrix of the full system as $\rho_{S_1S_2}$ and its
deviation as $\sigma_{S_1S_2}$, and use a similar notation with
subscripts $S_1$ and $S_2$ for the two subsystems. If the
evolution of system
 is governed by the ``noninteracting''
Hamiltonian of the form $H_{S_1S_2}=H_{S_1}+H_{S_2}$, where the terms
$H_{S_1}$, $H_{S_2}$ act only on variables of the system $S_1$,
$S_2$, respectively, then the overall norm $D_{S_1S_2}$ can
 be bounded by the sum of the norms $D_{S_1}$ and $D_{S_2}$,
\begin{eqnarray}\label{D12}
&&D_{S_1S_2}^{\vphantom{(i)}} = \sup_{\rho (0)} \| \sigma_{S_1S_2}^{\vphantom{(i)}} \|_{\lambda}
  = \sup_{\rho (0)} \| \rho_{S_1S_2}^{\vphantom{(i)}} - \rho^{(i)}_{S_1
S_2}\|_{\lambda}=\nonumber  \\
&&\sup_{\rho (0)}\| \rho_{S_1}^{\vphantom{(i)}} \! {\raise2pt\hbox{$\scriptscriptstyle{\otimes}$}}  \rho_{S_2}^{\vphantom{(i)}} \! - \rho_{S_1}^{(i)} {\raise2pt\hbox{$\scriptscriptstyle{\otimes}$}}
\rho_{S_2}^{(i)}\|_{\lambda}
  = \sup_{\rho (0)}\| \sigma_{S_1}^{\vphantom{(i)}} \! {\raise2pt\hbox{$\scriptscriptstyle{\otimes}$}}    \rho_{S_2}^{\vphantom{(i)}} \! + \rho_{S_1}^{(i)} {\raise2pt\hbox{$\scriptscriptstyle{\otimes}$}}
\sigma_{S_2}^{\vphantom{(i)}} \|_{\lambda}\nonumber\\
  &&\leq \sup_{\rho (0)} \| \sigma_{S_1}^{\vphantom{(i)}} {\raise2pt\hbox{$\scriptscriptstyle{\otimes}$}}  \rho_{S_2}^{\vphantom{(i)}} \|_{\lambda}
  +  \sup_{\rho (0)} \|  \rho_{S_1} ^{(i)}
{\raise2pt\hbox{$\scriptscriptstyle{\otimes}$}} \sigma_{S_2}^{\vphantom{(i)}} \|_{\lambda}
\nonumber\\
&&\leq\sup_{\rho_{S_1} (0)}\| \sigma_{S_1}^{\vphantom{(i)}} \|_{\lambda} +
\sup_{\rho_{S_2} (0)}\| \sigma_{S_2}^{\vphantom{(i)}}\|_{\lambda} =  D_{S_1}^{\vphantom{(i)}} +
D_{S_2}^{\vphantom{(i)}}.
\end{eqnarray}\par\noindent

In general, initially unentangled qubits will remain nearly
unentangled for short times, because they didn't have enough time
to interact. Therefore, the inequality $D_{S_1S_2}\lesssim
D_{S_1}+D_{S_2}$ is expected to provide a good approximate
estimate for $D_{S_1S_2}$, i.e., the measures for individual
qubits can be considered \emph{approximately\/} additive. For
large times, the separate measures become of order 1, so this
bound is not useful. Instead, the \emph{rates\/} of approach of
various quantities to their asymptotic values are approximately
additive in some cases. In the rest of this work, we focus on the
short-time regime, of no significant interaction effects, and
establish a much stronger property: We prove the approximate
additivity for initially \emph{entangled\/} qubits whose dynamics
is governed by (\ref{1}-\ref{3}) with coupling to independent
baths.

For intermediate calculations, we use the ``diamond'' norm
introduced by
Kitaev \cite{Kitaev}, which we denote by $K$,
\begin{equation}\label{supernormK}
K(t) =\|T- T^{(i)}\|_{\diamond}=\underset{ {\varrho}}{\sup}\|
\{[T-T^{(i)}]{\raise2pt\hbox{$\scriptscriptstyle{\otimes}$}} I_{G}\} {\varrho} \|_{\rm Tr}.
\end{equation}
Here, the trace norm \cite{Kato} is given by $\left\| A
\right\|_{\rm
Tr}={\rm Tr}\,\sqrt { A^{\dagger} A}$. The two
superoperators are defined as generating the actual and ``ideal''
dynamical evolutions,
\begin{equation}\label{R}
    T(t): \rho(0)\mapsto\rho(t),\quad  T^{(i)}(t):
\rho(0)\mapsto\rho^{(i)}(t),
\end{equation}
$I_{G}$ is the identity
superoperator on a Hilbert space $G$ whose dimension is the same
as that of the corresponding space of the superoperators $T$ and
$T^{(i)}$,
and $\varrho$ is a density operator in this space of twice the
number of qubits.

Consider again the composite system consisting of the
two subsystems $S_1$, $S_2$, with the
``noninteracting'' Hamiltonian $H_{S_1S_2}=H_{S_1}+H_{S_2}$. The
evolution superoperator of the system
will be $T_{S_1S_2}=T_{S_1}{\raise2pt\hbox{$\scriptscriptstyle{\otimes}$}} T_{S_2}$, and the ideal one
will
be $T_{S_1S_2}^{(i)}=T_{S_1}^{(i)}{\raise2pt\hbox{$\scriptscriptstyle{\otimes}$}} T_{S_2}^{(i)}$. The
diamond measure for the whole system can be expressed
\cite{Kitaev} as
\begin{eqnarray}
&&K_{S_1S_2}^{\vphantom{(i)}}=\|T_{S_1S_2}^{\vphantom{(i)}} -
T_{S_1S_2}^{(i)}\|_{\diamond}=\nonumber\\
&& \|(T_{S_1}^{\vphantom{(i)}}-T_{S_1}^{(i)}){\raise2pt\hbox{$\scriptscriptstyle{\otimes}$}} T_{S_2}^{\vphantom{(i)}}+T_{S_1}^{(i)}{\raise2pt\hbox{$\scriptscriptstyle{\otimes}$}}
(T_{S_2}^{\vphantom{(i)}}-T_{S_2}^{(i)})\|_{\diamond}\nonumber\\
&&\leq\|(T_{S_1}^{\vphantom{(i)}}-T_{S_1}^{(i)}){\raise2pt\hbox{$\scriptscriptstyle{\otimes}$}}
T_{S_2}^{\vphantom{(i)}}\|_{\diamond}+\|T_{S_1}^{(i)}{\raise2pt\hbox{$\scriptscriptstyle{\otimes}$}}
(T_{S_2}^{\vphantom{(i)}}-T_{S_2}^{(i)})\|_{\diamond}=\nonumber\\
&& \|T_{S_1}^{\vphantom{(i)}}-T_{S_1}^{(i)}\|_{\diamond}\| T_{S_2}^{\vphantom{(i)}}\|_{\diamond}
+\|T_{S_1}^{(i)}\|_{\diamond}\|T_{S_2}^{\vphantom{(i)}}-
T_{S_2}^{(i)}\|_{\diamond}=\nonumber\\
&& \|T_{S_1}^{\vphantom{(i)}}-T_{S_1}^{(i)}\|_{\diamond}+\|T_{S_2}^{\vphantom{(i)}}-
T_{S_2}^{(i)}\|_{\diamond}=K_{S_1}^{\vphantom{(i)}}+K_{S_2}^{\vphantom{(i)}}.\label{justbelow}
\end{eqnarray}
The step that follows the inequality, in (\ref{justbelow}), 
proved in \cite{Kitaev} as the ``stability'' property, was established
specifically with the definition (\ref{supernormK}). 

The approximate inequality $K_{S_1S_2}\lesssim K_{S_1}+K_{S_2}$
for the diamond norm $K$ has thus the same form as for the norm
$D$. Let us emphasize that both relations apply assuming that for
short times the subsystem interactions, directly with each other
or via their coupling to the bath modes, have had no significant
effect. However, there is an important difference that the
relation for $D$ further requires the subsystems to be initially
unentangled. This restriction does not apply to the relation for
$K$. This property is particularly useful for quantum computing, 
which is based on qubit entanglement.
However, even in the simplest case of the diamond norm of one
qubit, the calculations are extremely cumbersome. Therefore, the
measure $D$ is favorable for actual calculations.

In general, one can prove that $ 2 D(t) \le
K(t)\le 2$. For a single qubit, calculation within
the short-time approximation actually gives $K(t)=2 D(t)$. Since
$D$ is generally bounded by $K/2$, it follows that for the
specific model considered, with a bosonic heat bath acting
independently on each of the qubits, the multiqubit norm $D$ is
approximately bounded from above by the sum of the single-qubit
norms even for the \emph{initially entangled\/} qubits:
\begin{equation}\label{DN1}
    D \le K/2 \lesssim \hbox{$\sum_j$}K_{j}/2=\hbox{$\sum_j$}(1 -
e^{ -
B^2_j})/2 = \hbox{$\sum_j D_{j}$},
\end{equation}
where $j = 1, \ldots, N$ labels the qubits.

For short times, one can also establish a lower bound on $D(t)$.
Consider a
specific initial state with all the qubits excited,
$\rho(0)=(|2\rangle\langle 2|)_1 {\raise2pt\hbox{$\scriptscriptstyle{\otimes}$}} \ldots {\raise2pt\hbox{$\scriptscriptstyle{\otimes}$}}
(|2\rangle\langle
2|)_N$. Then according to (\ref{short-time}),
$ \rho(t) = \rho_1 (t) {{\raise2pt\hbox{$\scriptscriptstyle{\otimes}$}}}
\ldots {\raise2pt\hbox{$\scriptscriptstyle{\otimes}$}} \rho_N(t)$, where
 $\rho_j (t)=(1/2) \big\{\big[1-e^{-
B^2_j(t)}\big]|1 \rangle\langle 1|+ \big[1+e^{-B^2_j(t)}\big]|2
\rangle\langle 2|\big\}$. The right-bottom matrix element of the
(diagonal) deviation operator,
$\sigma_{2^N,2^N}(t)=-1+2^{-N}{\prod_j}[1+e^{-B^2_j(t)}]$, can be
expanded as $({1}/{2})\sum_j B_j^2 (t)+o(\sum_j B_j^2 (t))$,
because $B(t)$ is linear in $t$ for small times. The largest
eigenvalue of $\sigma (t)$ cannot be smaller than
$\sigma_{2^N,2^N}(t)$. It follows that
\begin{equation}\label{DN0}
  D\geq \hbox{$\frac 1 2$}\hbox{$\sum_j$} B^2_j
+o\big(\hbox{$\sum_j$} B_j^2 \big)= \hbox{$\sum_j
$}D_{j}+o\big(\hbox{$\sum_j $} D_{j}\big),
\end{equation}
where we used (\ref{D1qubit}) for short times. 

By combining the
upper and lower bounds, we get the final result for short times,
\begin{equation}\label{DN4}
    D(t)=\hbox{$\sum_j$} D_{j}(t)+o\big(\hbox{$\sum_j$}
D_{j}(t)\big).
\end{equation}
This completes the proof of additivity of the measure $D(t)$ at
short times, for
spin-qubits interacting with bosonic environmental modes.
This result should apply as a good approximation up to intermediate, inverse-system-energy-gap times, e.g., \cite{short}, and it is consistent with the recent finding \cite{Dalton} that at short times decoherence
of a trapped-ion quantum computer scales approximately linearly
with the number of qubits.

\begin{acknowledgments}
This research was supported by the NSA and
ARDA under ARO contract DAAD-190210035, and by the NSF, grant DMR-0121146.
\end{acknowledgments}


\begin{thebibliography}{15}

\bibitem{open}
G.W.\ Ford, M.\ Kac and P.\ Mazur, J.\ Math.\ Phys.\ \textbf{6}, 504
(1965);  A.O.\ Caldeira and A.J.\ Leggett, Physica A \textbf{121},
587 (1983); S.\ Chakravarty and A.J.\ Leggett, Phys.\ Rev.\ Lett.\
\textbf{52}, 5 (1984); A.J.\ Leggett, S.\ Chakravarty, A.T.\ Dorsey,
M.P.A.\ Fisher and W.\ Zwerger, Rev.\ Mod.\ Phys.\ \textbf{59}, 1
(1987); H.\ Grabert, P.\ Schramm and G.-L.\ Ingold, Phys.\ Rep.\
\textbf{168}, 115 (1988).

\bibitem{vanKampen}
N.G.\ van Kampen, J.\ Stat.\ Phys.\ \textbf{78}, 299 (1995).

\bibitem{nonMarkov}
K.M.\ Fonseca Romero and M.C.\ Nemes, Phys.\ Lett.\ A \textbf{235}, 432
(1997); C.\ Anastopoulos and B.L.\ Hu, Phys.\ Rev.\ A \textbf{62},
033821 (2000); G.W.\ Ford and R.F.\ O'Connell, Phys.\ Rev.\ D
\textbf{64}, 105020 (2001); D.\ Braun, F.\ Haake, and W.T.\ Strunz,
Phys.\ Rev.\ Lett.\ \textbf{86}, 2913 (2001); G.W.\ Ford, J.T.\ Lewis,
and R.F.\ O'Connell, Phys.\ Rev.\ A \textbf{64}, 032101 (2001); J.\
Wang, H.E.\ Ruda and B.\ Qiao, Phys.\ Lett.\ A \textbf{294}, 6
(2002); E.\ Lutz, cond-mat/0208503; A.\ Khaetskii, D.\ Loss and L.\
Glazman, Phys.\ Rev.\ B \textbf{67}, 195329 (2003);  R.F.\ O'Connell
and J.\ Zuo, Phys.\ Rev.\ A \textbf{67}, 062107 (2003), W.T.\ Strunz,
F.\ Haake and D.\ Braun, Phys.\ Rev.\ A \textbf{67}, 022101 (2003);
W.T.\ Strunz and F.\ Haake, Phys.\ Rev.\ A  \textbf{67}, 022102
(2003); V.\ Privman, D.\ Mozyrsky and I.D.\ Vagner, Comp.\ Phys.\ Commun.\ \textbf{146}, 331 (2002).

\bibitem{short}
V.\ Privman, J.\ Stat.\ Phys.\ \textbf{110}, 957 (2003).

\bibitem{Privman}
V.\ Privman, Mod.\ Phys.\ Lett.\ B \textbf{16}, 459 (2002).

\bibitem{qec} P.W.\ Shor, Phys.\
Rev.\ A {\bf 52},  R2493 (1995); A.M.\ Steane, Phys.\ Rev.\ Lett.\ {\bf
77}, 793 (1996); C.H.\ Bennett, G.\ Brassard, S.\ Popescu,
B.\ Schumacher, J.A.\ Smolin and W.K.\ Wootters, Phys.\ Rev.\ Lett.\
{\bf 76}, 722 (1996); A.R.\ Calderbank and P.W.\ Shor, Phys.\ Rev.\ A
{\bf 54}, 1098 (1996); A.M.\ Steane, Phys.\ Rev.\ A {\bf 54}, 4741
(1996); D.\ Aharonov and M.\ Ben-Or, quant-ph/9611025; D.\ Gottesman,
Phys.\ Rev.\ A {\bf 54}, 1862 (1997); E.\ Knill and R.\ Laflamme,
Phys.\ Rev.\ A {\bf 55}, 900 (1997).

\bibitem{apscheme}
D.\ Loss and D.P.\ DiVincenzo, cond-mat/0304118; V.\ Privman, Proc.\
SPIE \textbf{5115}, 345 (2003); W.H.\ Zurek, Rev.\ Mod.\ Phys.\ \textbf{75}, 715 (2003).

\bibitem{Caldeira}
A.O.\ Caldeira and A.J.\ Leggett, Phys.\ Rev.\ Lett.\ \textbf{46}, 211
(1981).

\bibitem{Lloyd}
S.\ Lloyd, Phys.\ Rev.\ Lett.\ \textbf{75}, 346 (1995).

\bibitem{Barenco}
A.\ Barenco, C.H.\ Bennett, R.\ Cleve, D.P.\ DiVincenzo, N.\ Margolus,
P.\ Shor, T.\ Sleator, J.A.\ Smolin and H.\ Weinfurter, Phys.\ Rev.\ A
\textbf{52}, 3457 (1995).

\bibitem{Markov}
N.G.\ van Kampen, \emph{Stochastic Processes in Physics and
Chemistry\/}, North-Holland, 2001; W.H.\ Louisell, \emph{Quantum
Statistical Properties of Radiation\/}, Wiley, 1973; K.\ Blum,
\emph{Density Matrix Theory and Applications\/}, Plenum, 1996; A.\
Abragam, \emph{The Principles of Nuclear Magnetism\/}, Clarendon
Press, 1983.

\bibitem{Kitaev}
A.Y.\ Kitaev, \emph{Russ.\ Math.\ Surv.\/} \textbf{52}, 1191 (1997);
D.\ Aharonov, A.\ Kitaev and N.\ Nisan, Proc.\ XXXth ACM Symp.\ Theor.\ Comp., Dallas, TX, USA, 20 (1998); A.Yu.\ Kitaev,
A.H.\ Shen and M.N.\ Vyalyi, \emph{Classical and Quantum
Computation\/}, AMS, 2002.

\bibitem{PreskillDiVincenzo} J.\ Preskill, Proc.\ Roy.\ Soc.\ Lond.\ A
{\bf 454}, 385 (1998); 
D.P.\ DiVincenzo, Fort.\ Phys.\ \textbf{48}, 771 (2000).

\bibitem{Storcz}
M.J.\ Storcz and F.K.\ Wilhelm, Phys.\ Rev.\ A \textbf{67}, 042319
(2003).

\bibitem{Palma}
G.M.\ Palma, K.A.\ Suominen and A.K.\ Ekert, Proc.\ Roy.\ Soc.\ Lond.\ A
\textbf{452}, 567 (1996).

\bibitem{DFS}
L.-M.\ Duan and G.-C.\ Guo, Phys.\ Rev.\ Lett.\ \textbf{79}, 1953
(1997); P.\ Zanardi and M.\ Rasetti, Phys.\ Rev.\ Lett.\ \textbf{79},
3306 (1997); D.A.\ Lidar, I.L.\ Chuang and K.B.\ Whaley, Phys.\ Rev.\
Lett.\ \textbf{81}, 2594 (1998).

\bibitem{norm}
L.\ Fedichkin, A.\ Fedorov and V.\ Privman, Proc.\ SPIE
\textbf{5105}, 243 (2003).

\bibitem{Kato}
T. Kato, \emph{Perturbation Theory for Linear Operators\/},
Springer-Verlag, NY, 1995.


\bibitem{basis}
D.\ Mozyrsky and V.\ Privman, J.\ Stat.\ Phys.\ \textbf{91}, 787
(1998).

\bibitem{Dalton}
B.J.\ Dalton, J.\ Mod.\ Opt.\ \textbf{50}, 951 (2003).

\end{thebibliography}
\end{document}